\documentclass[vecphys]{svmult}

\usepackage[utf8]{inputenc}

\usepackage{makeidx}         
\usepackage{graphicx}        
\graphicspath{{figures/}{figures/intro/}{figures/theory/}{figures/results/}}

\usepackage{multicol}        
\usepackage{cite}            
\usepackage[bottom]{footmisc}

\makeindex             

\usepackage{amsmath}
\usepackage{amsfonts}
\usepackage{amssymb}
\usepackage{amstext, mathrsfs, textcomp}
\usepackage{xcolor}

\usepackage{layouts}

\begin{document}

\title{Fano-Resonances in High Index Dielectric Nanowires for Directional Scattering}
\titlerunning{Fano-Resonances for Directional Scattering in Dielectric Nanowires}

\author{Peter R. Wiecha$^1$ \and
	Aur\'elien Cuche$^1$ \and
	Houssem Kallel$^1$ \and
	G\'erard Colas des Francs$^2$ \and
	Aur\'elie Lecestre$^3$ \and
	Guilhem Larrieu$^3$ \and
	Vincent Larrey$^{4}$ \and
	Frank Fournel$^{4}$ \and
	Thierry Baron$^{5}$ \and
	Arnaud Arbouet$^1$ \and
	Vincent Paillard$^1$}

\authorrunning{Peter R. Wiecha \textit{et al.}}

\institute{CEMES, Universit\'e de Toulouse, CNRS, Toulouse, France\\
\texttt{peter.wiecha@cemes.fr}, \texttt{vincent.paillard@cemes.fr}
\and ICB, Universit\'e de Bourgogne-Franche Comt\'e, CNRS, Dijon, France
\and LAAS, Universit\'e de Toulouse, CNRS, Toulouse, France
\and CEA-LETI, Universit\'e Grenoble-Alpes, MINATEC Campus, Grenoble, France
\and LTM, Universit\'e Grenoble-Alpes, CNRS, Grenoble, France}

\maketitle

\begin{abstract}
High refractive index dielectric nanostructures provide original optical properties thanks to the occurrence of size- and shape-dependent optical resonance modes. 
These modes commonly present a spectral overlap of broad, low-order modes (\textit{e.g}. dipolar modes) and much narrower, higher-order modes. 
The latter are usually characterized by a rapidly varying frequency-dependent phase, which -- in superposition with the lower order mode of approximately constant phase -- leads to typical spectral features known as Fano resonances.
Interestingly, such Fano resonances occur in dielectric nanostructures of the simplest shapes. 
In spheroidal nanoparticles, interference between broad magnetic dipole and narrower electric dipole modes can be observed. 
In high aspect-ratio structures like nanowires, either the electric or the magnetic dipolar mode (depending on the illumination conditions) interferes with higher order multipole contributions of the same nature (electric or magnetic).
Using the analytical Mie theory, we analyze the occurrence of Fano resonances in high-index dielectric nanowires and discuss their consequences like unidirectional scattering. 
By means of numerical simulations, we furthermore study the impact on those Fano resonances of the shape of the nanowire cross-sections as well as the coupling of two parallel nanowires.
The presented results show that all-dielectric nanostructures, even of simple shapes, provide a reliable low-loss alternative to plasmonic nanoantennas.
\end{abstract}

\noindent

\clearpage
\section{Introduction}

Research on effects of light-matter interaction occurring at subwavelength dimensions   has been drawing increasing attention during the last three decades.
When we talk about subwavelength dimensions, at visible and near-infrared frequencies, we find ourselves at the nanometer scale, a length-scale particularly interesting with regards to applications in information processing and optical computing, single molecule sensing or biomedicine, amongst many other domains.
Most applications rely on the possibility to strongly confine far-field radiation to deeply subwavelength small volumes at resonant modes of specifically designed nanoparticles. Resonances are either due to surface plasmons in the case of metals (field of plasmonics) or to constructive optical interference in the case of dielectrics.
Usually, all phenomena can be described by classical electrodynamics, \emph{i.e.} by the set of Maxwell's equations \cite{maxwell_dynamical_1865}.

In the following, we will briefly describe the fields of plasmonic and high index dielectric nanostructures, with a particular attention on their specificity and main differences.

Then, in the other sections of this chapter, we will give an overview on Fano Resonances in dielectric nanostructures, with an emphasis on directional scattering as a result of Fano-like interference phenomena.
In particular, we will present the so-called \textit{Kerker's conditions} under which unidirectional scattering occurs in dielectric spherical particles and extend the idea to cylinders.
To substantiate the findings on dielectric nanowires, we will compare experimental results to Mie theory and numerical simulations.

Finally, we analyze the directional scattering behavior of nanowires with nonsymmetric cross-sections, as well as the case of a system of two coupled nanowires.

\subsection{Plasmonics}
  
  One of the main driving forces in nano-optics is the field of \textit{plasmonics}\index{Plasmonics} \cite{muhlschlegel_resonant_2005, maier_plasmonics_2010}.
  Electromagnetic waves impinging on metals launch collective oscillations of the 
  free electrons in the conduction band of the metal.
  The dielectric constant of metals is negative, leading to an imaginary wave vector. 
  Fields are therefore evanescent and confined within a small region at the surface, called the \textit{skin-depth}\index{Skin depth}.
  In consequence, the collective oscillations of the electrons propagate along the surface and are therefore called \textit{surface plasmon polaritons} (SPP)\index{Surface plasmon polaritons}. 
  In small metallic particles, the propagation of SPPs is hindered due to the spatial confinement and localized modes appear, so-called \textit{localized surface plasmon resonances} (LSP)\index{Localized surface plasmon resonances}.
  
  These confined plasmon oscillations allow to squeeze light into tiny volumes of subwavelength size, far below the diffraction limit and yield extremely high local field enhancements \cite{schuck_improving_2005}. 
  In the visible spectral range, this results in sizes of several tenths to a few hundreds of nanometers for resonant metallic nanostructures. 
  Such plasmonic particles are often referred to as \textit{optical (nano-)antennas}\index{Optical antennas} \cite{muhlschlegel_resonant_2005}. 

  In the context on this chapter, we would like to mention a few selected examples for applications of plasmonic nanostructures.
  Beyond the possibility to obtain strong localized field intensities, plasmonic nanoantennas can be designed to provide directional scattering.
  This can be achieved for instance by tailoring plasmonic geometries which provide a simultaneous electric and magnetic response \cite{yao_controlling_2016}.
  With complex structures like bimetallic antennas, wavelength selective directional color routing can be obtained from individual nanoantennas \cite{shegai_bimetallic_2011}.
  Also the directional emission of quantum emitters can be controlled using plasmonic antennas \cite{bonod_ultracompact_2010, curto_unidirectional_2010, hancu_multipolar_2014}.
  For an extensive introduction on plasmonics, see \textit{e.g.} reference~\cite{maier_plasmonics_2010}.

\subsection{High Refractive Index Dielectric Nano-Particles}

The focus of this chapter lies on a different kind of nano-antennas than the plasmonic ones.
Recently, \textit{dielectric} nanostructures of high refractive index started drawing a lot of attention as promising alternatives to metallic particles, since they offer in a similar way tailorable optical resonances  \cite{kuznetsov_optically_2016}.

Even in very simple systems such as nanospheres \cite{kuznetsov_magnetic_2012}, or cylindrical \cite{cao_tuning_2010, kallel_tunable_2012} or rectangular nanowires (NWs) \cite{ee_shape-dependent_2015}, optical resonance modes can be tuned over (and beyond) the whole visible spectral range.
Flexible tailoring of the resonant behavior can be achieved using dielectric nanostructures of more complex shapes \cite{wiecha_evolutionary_2017}.

\begin{figure}[t]
  \centering
  \includegraphics*{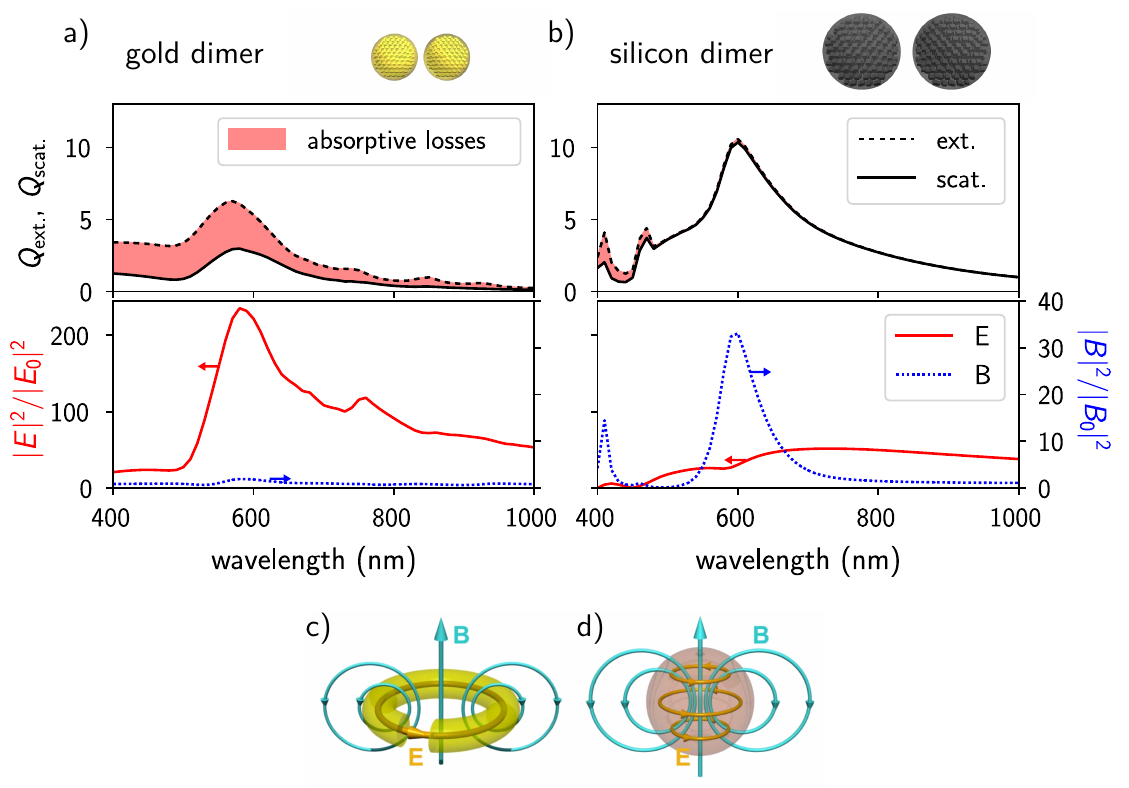}
  \caption[Plasmonic vs. dielectric nanostructures]{
  Extinction and scattering efficiency spectra (top) as well as nearfield enhancement (bottom: \(\mathbf{E}\)-field: red, \(\mathbf{B}\)-field: blue) in the gap of
  (a) a dimer of two silicon spheres with radius \(R=80\)\,nm and 
  (b) a gold dimer of nano-spheres with \(R=40\)\,nm. 
  Gap is \(R/4\) in both cases. 
  Sizes were chosen to obtain resonances around \(\lambda=600\,\)nm. 
  Incident plane wave polarized along dimer axis. 
  Dimers are placed in vacuum.
  A sketch of the model and numerical mesh used in the simulations for the gold and silicon dimers is shown above the plots.
  (c-d) illustration of the mechanism for obtaining strong magnetic fields in (c) plasmonics using a circular current density in a metal nano-ring and (d) in dielectrics, occurring naturally via the curl of the electric displacement current in geometries as simple as spheres (c-d reprinted by permission from Macmillan Publishers Ltd from \cite{kuznetsov_magnetic_2012}, copyright 2012)}
  \label{fig:plasmonic_vs_dielectric}
\end{figure}

A particular advantage of dielectric particles over metallic ones are their very low losses \cite{albella_electric_2014, caldarola_non-plasmonic_2015, decker_resonant_2016}. 
This property is directly related to the generally low imaginary part of the dielectric function for wavelengths above the direct bandgap. It is demonstrated in the upper spectra of figure~\ref{fig:plasmonic_vs_dielectric}(a-b), where a gold and a silicon dimer are compared in terms of scattering efficiency and dissipation.
Although the electric field-enhancement (as well as the confinement) is usually at least an order of magnitude lower than for plasmonic antennas (see also Fig.~\ref{fig:plasmonic_vs_dielectric}(a-b), lower spectra), the reduced dissipative losses are a tremendous advantage and can be a decisive factor in applications such as field enhanced spectroscopy \cite{wells_silicon_2012, regmi_all-dielectric_2016}.

Another unique feature of dielectric nanoparticles is the possibility to obtain strong magnetic resonances from geometries as simple as a sphere \cite{garcia-etxarri_strong_2011, ginn_realizing_2012, evlyukhin_demonstration_2012, valuckas_direct_2017}.\index{Magnetic resonances}
In contrast, plasmonics require complex geometries to obtain important magnetic resonances \cite{kuznetsov_magnetic_2012} (see figure~\ref{fig:plasmonic_vs_dielectric}c and~\ref{fig:plasmonic_vs_dielectric}d).
These magnetic-type resonances allow also to obtain a strong enhancement of the magnetic near-field, which is usually significantly more intense than the magnetic field intensity obtainable using metal nanostructures \cite{albella_electric_2014, bakker_magnetic_2015, mirzaei_electric_2015} (see figure~\ref{fig:plasmonic_vs_dielectric}(a-b), bottom spectra).
The magnetic resonances of dielectric nanostructures can be exploited to enhance the decay rate of magnetic dipole transitions \cite{rolly_promoting_2012, schmidt_dielectric_2012, baranov_modifying_2017, wiecha_decay_2017}.
While the optical magnetic near-field and the magnetic contribution to the local density of photonic states (LDOS)\index{Local density of photonic states} in the vicinity of nanostructures can be probed using appropriate SNOM tips (gold-ring coated tips for sensing the \(\mathbf{B}\)-field \cite{devaux_local_2000, burresi_probing_2009}; tips prepared with rare-earth-ion doped nano-crystals, e.g. using Eu\(^{3+}\), for the magnetic LDOS \cite{ aigouy_mapping_2014, carminati_electromagnetic_2015}), it is experimentally far more demanding to access the intrinsic field enhancement.
While no measurement was reported in the visible regime, strong magnetic field enhancement in dielectric cylinders has recently been experimentally demonstrated at THZ frequencies \cite{kapitanova_giant_2017}.

\subsection{General Applications of Dielectric Nano-Structures}

High-index dielectric nanostructures are increasingly used in many nano-optical applications, often -- but not exclusively -- as alternatives to their lossy plasmonic equivalent.

In perfect analogy to plasmonics, it is possible to design color filters \cite{wood_all-dielectric_2017} and (color) holograms \cite{zhao_full-color_2016, li_dielectric_2017-1} or to use dielectric nano-structures for ``color-printing'' at the diffraction limit \cite{proust_all-dielectric_2016, flauraud_silicon_2017}.

Field-enhanced spectroscopy can even more benefit by replacing plasmonic-based substrates with dielectric nanoantenna-based substrates for two main reasons.
The first reason is the prevention of the reabsorption of the generated fluorescence or Raman signal that occurs in the metal nanostructures \cite{gerard_strong_2008, wells_silicon_2012, caldarola_non-plasmonic_2015, regmi_all-dielectric_2016, cambiasso_bridging_2017}.
The second reason is  that heat generation can be almost completely suppressed using dielectric nanoantennas, while plasmonic structures suffer from strong local heating \cite{albella_low-loss_2013, albella_electric_2014}. This point is important to realize very sensitive biosensors \cite{bontempi_highly_2017}, as a slight increase of temperature can be fatal for fragile biomolecules or simply decrease the luminescence intensity.

Further applications of photonic nano-particles based on dielectric materials can be found in photovoltaics: The geometrical structure of the photovoltaic junction can be shaped such that its absorption covers optimally the solar spectrum.
It has been shown that already simple geometries like nano-blocks or nanowires can significantly improve the absorptive coverage of the solar spectrum, compared to two-dimensional layers used in commercial photovoltaic cells \cite{cao_semiconductor_2010, kallel_enhanced_2013}. 
A different approach is to design dielectric (low-loss) nano-antennas able to trap and redirect the incoming solar light towards the classical (planar) photovoltaic junction \cite{brongersma_light_2014, priolo_silicon_2014}.

Finally, the strong field enhancements occurring at the resonant modes of dielectric nanostructures can be used to promote nonlinear optical effects. 
For instance, it has been shown that surface second harmonic generation (SHG) can be strongly enhanced in dielectric nanoparticles \cite{wiecha_enhanced_2015, wiecha_origin_2016, liu_resonantly_2016, cambiasso_bridging_2017}. 
Also the third harmonic generation (THG) can be significantly enhanced \cite{shcherbakov_enhanced_2014, melik-gaykazyan_third-harmonic_2017} and its emission can be tailored \cite{shcherbakov_nonlinear_2015, wang_shaping_2017} in silicon nano-particles, by making use of their magnetic resonances.
Even ultrafast all-optical switching of the optical transmission has been demonstrated using silicon nano-discs \cite{shcherbakov_ultrafast_2015}.

\section{Fano Resonances and Kerker's Conditions}

\begin{figure}[t]
  \centering
  \includegraphics*{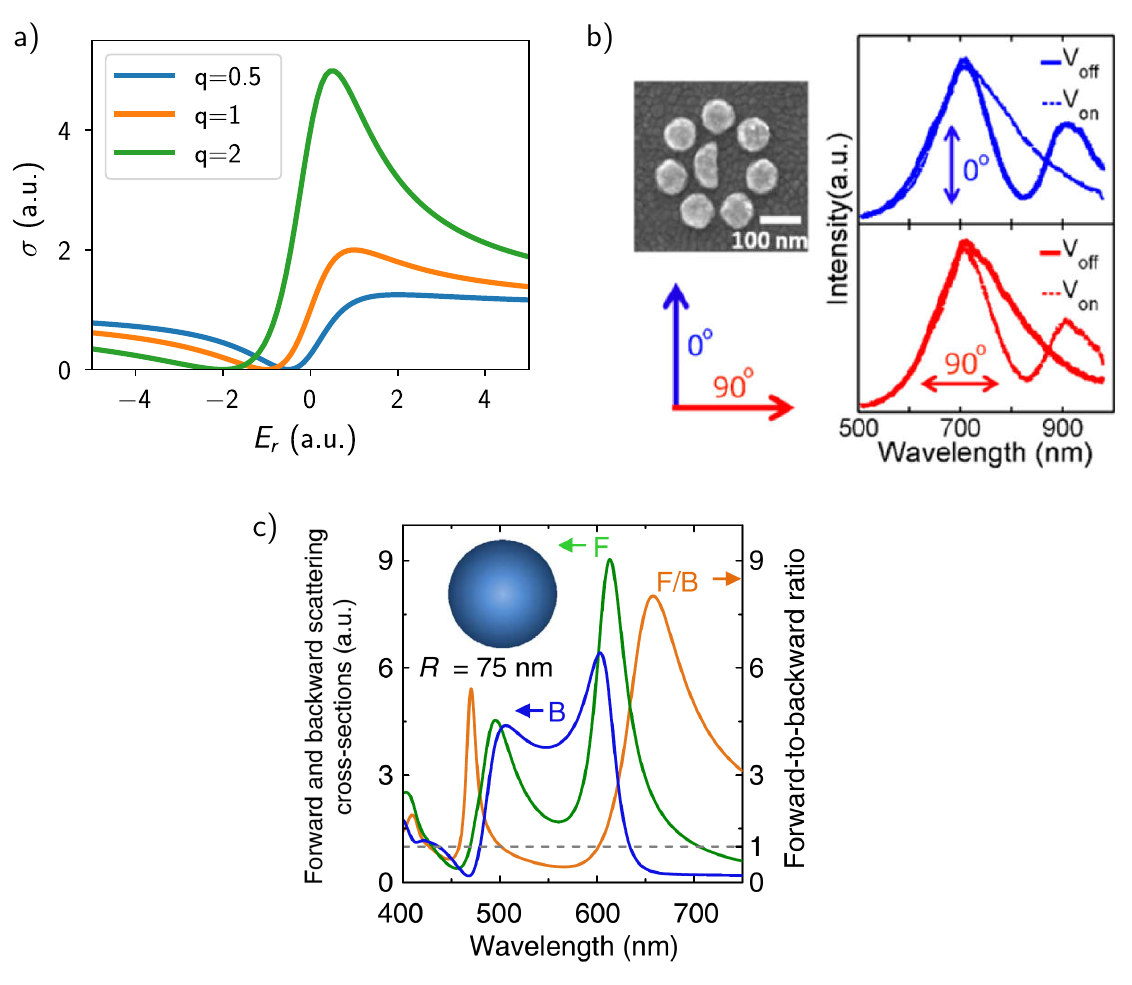}
  \caption[Optical Fano resonances in nano-structures]{
  a) Fano profiles for Fano parameters \(q=0.5\), \(q=1\) and \(q=2\). 
  b) Plasmonic ``Fano switch'', which allows to toggle the polarization dependent transmittance of a metallic structure embedded in a liquid crystal via a switchable Fano resonance, reprinted with permission from \cite{chang_plasmonic_2012}. Copyright 2012 American Chemical Society.
  c) Directional resolved scattering spectra from a silicon nano-sphere of radius \(R=75\,\)nm, reprinted by permission from Macmillan Publishers Ltd from \cite{fu_directional_2013}, copyright 2013
  } 
  \label{fig:fano}
\end{figure}

Fano resonances\index{Fano resonance} occur due to the interference between two scattering amplitudes, when a resonant state energetically lies in a background of continuous states.
If the scattering amplitudes of the resonance and the background are of comparable magnitude, the cross section of the Fano resonance follows a very characteristic, asymmetric lineshape. 
Although Ugo Fano originally considered the interference between a single state with a continuum \cite{fano_effects_1961}, very similar resonance profiles occur if a narrow resonant state interferes with a significantly broader state. 
Such interference between two resonant states of similar amplitude but different linewidths are nowadays often colloquially called \textit{Fano resonances}.
The Fano lineshape can be written as
\begin{equation}\label{eq:fano_resonance}
 \sigma(E_r) \propto 1 + \frac{q^2 + 2q E_r - 1}{ 1 + E_r^2}
\end{equation}
with the reduced energy 
\begin{equation}
 E_r = \frac{E - E_0}{\Gamma/2},
\end{equation}
where \(E_0\) is the position of the resonance and \(\Gamma\) it's spectral width.
The amplitude of the background state(s) is considered constant over the spectral range of the Fano profile.
\(q\) is called the \textit{Fano parameter}\index{Fano parameter}, which corresponds to the ratio between the resonance amplitude and the non-resonant background.
Note that in the limit of a very strong resonance on a weak background, equation~\eqref{eq:fano_resonance} converges towards a Lorentzian.
On the other hand, for \(q\approx 1\) a strong interference between the resonance and the background exists, which results in the typical line shapes.
The cases \(q=0.5\), \(q=1\), and \(q=2\) are shown in figure~\ref{fig:fano}a.

For a detailed coupled mode theory of Fano resonances in optical resonators, see also reference~\cite{fan_temporal_2003}.

\subsection{Fano in nano-optics}
Optical Fano resonances in photonic nanostructures allow to obtain anomalous, usually very sharp mode signatures in the scattering or extinction spectra.
The most prominent structure geometry which pronounces Fano-like mode profiles is the class of oligomers: Ordered arrangements of several nano-particles.
Tailorable Fano resonances have been observed for example in dielectric oligomers. 
The resonances occur as a result of an interference between the magnetic resonance of each individual constitute and a collective magnetic response of the entire system and can be tailored by the size and distance between the individual particles \cite{hopkins_interplay_2015}.
Also in plasmonic oligomers Fano resonances occur, for instance due to interference between electric and magnetic modes \cite{bakhti_fano-like_2016}.
Using an asymmetric arrangement of plasmonic particles, it is even possible to create an optical ``Fano-switch'' \cite{chang_plasmonic_2012}:
Via the polarization of the incident light, the Fano resonance can be toggled on and off, as shown in figure~\ref{fig:fano}b.

But Fano-like scattering profiles are not limited to complex geometrical arrangements like the above mentioned oligomers.
In fact, Fano resonances occur in very simple systems such as individual, dielectric nanowires of rectangular cross-section with relatively high aspect-ratio. 
Narrow Fano lineshapes appear in the scattering spectra of such nanowires, at which the scattering rapidly drops, while absorption strongly rises. 
This can be explained by a spectrally sharp guided mode along the NW width, which interferes with a spectrally broad leaky mode resonance \cite{fan_optical_2014}.
A similar kind of Fano resonance due to the interference between guided modes (along the NW axis) and Mie resonances has recently been described in short dielectric nanowires \cite{abujetas_high-contrast_2017}.
For a detailed review on Fano resonances in photonics, we would like to point to reference~\cite{limonov_fano_2017}.

\subsection{Kerker's Conditions at Optical Frequencies}
A special type of Fano resonance was studied by Kerker \textit{et al.} in hypothetical magneto-dielectric nanospheres, for which they found conditions at which exclusive forward (FW) or backward (BW) scattering occurs \cite{kerker_electromagnetic_1983}.
Kerker \textit{et al.} described two possible configurations, called the \textit{Kerker's conditions}. 
\index{Kerker's conditions}
The first Kerker's condition requires equal electric permittivity and magnetic permeability (\(\epsilon_r = \mu_r\)), at which zero-backward scattering occurs.
The second Kerker's condition predicts zero-forward scattering in the case if the first order magnetic Mie coefficient and first order electric Mie coefficient are of equal value but of opposite sign (\(a_1 = -b_1\)).
\index{First Kerker's condition}\index{Second Kerker's condition}

In contrast to particularly designed metamaterials \cite{pendry_controlling_2006}, no known material in nature has a direct response to rapidly oscillating magnetic fields.
Hence, \(\mu_r\) is unitary in dielectric nano-particles.
However, equally strong electric and magnetic resonances can \textit{de-facto} fulfill the first Kerker's condition, if they overlap energetically \cite{gomez-medina_electric_2011,nieto-vesperinas_angle-suppressed_2011}.
The second condition on the other hand contradicts the optical theorem and is therefore unphysical. 
Nevertheless, \textit{almost} zero-forward scattering can still be achieved, however scattering efficiencies are in that case usually considerably lower compared to the first Kerker's condition (zero-backward scattering) \cite{alu_how_2010, nieto-vesperinas_angle-suppressed_2011, wiecha_strongly_2017}. 

Note that, although the Kerker's conditions were originally derived for spherical particles, it has been shown later that they have their origin in a cylindrical symmetry and can hence be generalized to particles of cylindrical symmetry, excited along their long axis \cite{zambrana-puyalto_duality_2013}.


\subsection{Directional Scattering From Nanoparticles}

\subsubsection*{Directional Scattering from Nanospheres and Small Particles}

The possibility to \textit{de facto} fulfill the Kerker's conditions at visible wavelengths for unidirectional forward and backward scattering has been first discussed around 2010 for spherical high refractive index dielectric nanoparticles (silicon or germanium) \cite{nieto-vesperinas_angle-suppressed_2011, garcia-etxarri_strong_2011, gomez-medina_electric_2011}.
For instance, spectra showing the directional scattering of light from an individual silicon nanosphere are presented in figure~\ref{fig:fano}c.
It turned out that the directionality of light scattering occurs in many kinds of dielectric nanoparticles, like individual nanocubes or nanocube dimers \cite{campione_tailoring_2015}.
It has been shown as well, that the luminescence of emitting dipoles placed near the surface of spherical dielectric nanoparticles can be focused in a preferred direction via simultaneous excitation of electric and magnetic modes in the spheres \cite{rolly_boosting_2012}.
The same effect has been predicted for quantum emitters embedded inside dielectric nanodiscs\cite{rocco_controlling_2017}.
\index{Nano-spheres}
\index{Nano-spheroids}
\index{Nano-cubes}
\index{Nano-discs / -cylinders}

The experimental demonstration of Kerker-type scattering in the visible range was achieved about two years after the theoretical prediction.
In early 2013, two research groups around Novotny and Luk’yanchuk published almost at the same time their results on exclusive forward or backward scattering from dielectric nanoparticles \cite{person_demonstration_2013, fu_directional_2013}.
Since then, similar results have been  achieved for different geometries of dielectric nanoparticles, like nanodiscs \cite{staude_tailoring_2013}, nanospheroids \cite{lukyanchuk_optimum_2015} or nanosphere dimers for switchable directional scattering \cite{albella_switchable_2015, yan_directional_2015, shibanuma_experimental_2017}.

The main insight of all these works is that dielectric spheres permit a spectral overlap of electric and magnetic dipolar resonances, thus allowing to effectively satisfy Kerker's conditions and obtain a distinct directional scattering.

\subsubsection*{Directional Scattering from Nanowires}

In a nanosphere excited with a linearly polarized plane wave, transverse electric (TE) and transverse magnetic (TM) polarizations are not separable, hence the according modes are  always simultaneously excited.
In the case of normally illuminated nanowires by the plane wave\index{Nanowires} on the other hand, it is possible to excite either TE or TM modes by choosing an illumination with a polarization either perpendicular, or parallel, to the NW axis.
In that case, it becomes impossible to obtain spectrally overlapping electric and magnetic resonances, so that  Kerker's conditions (for instance \(a_1 = -b_1\)) cannot be satisfied.

Several propositions have been made to overcome this limitation and generate Kerker-type directional scattering from nanowires (NWs) as well. 
For instance, by using hybrid plasmonic / dielectric materials in core-shell NWs, an electric / magnetic response can be tailored, leading to directional scattering \cite{liu_scattering_2013} (a concept which, by the way, has been successfully applied also to ``non-wire'' nano-particles \cite{guo_multipolar_2016}).
In this context it has been demonstrated that properly designed plasmonic / dielectric core-shell NWs can be even rendered invisible thanks to destructive interference between electric and magnetic modes \cite{liu_invisible_2015}.
However, the implementation of plasmonic components would increase losses due to absorption in the metal, as well as the complexity of the object fabrication. In that regard, all-dielectric solutions seem advantageous.
In analogy to metal/dielectric hybrid NWs, Kerker-type directional scattering can occur in NWs with a radially anisotropic refractive index \cite{liu_superscattering_2017}.
Likewise, the invisibility effect can be achieved in multilayer all-dielectric NWs \cite{mirzaei_all-dielectric_2015}.

\subsubsection*{Directional Scattering from Complex Dielectric Nanostructures}

Obviously, complex geometries can be used to tailor the spectral positions of electric and magnetic resonances.
Without going in much detail, Let us refer to two examples as illustration.

First, it has been shown that V-shaped dielectric resonators can be designed for bidirectional color routing (in perpendicular directions with respect to the incidence). 
This directionality has been found to be a result of interference between electric dipolar, magnetic dipolar as well as an electric quadrupolar mode. 
Furthermore, the left/right color routing effect is observed in addition to the occurrence of an unidirectional forward / backward scattering \cite{li_all-dielectric_2016}.

As a second example for tailored scattering behavior, we would like to mention a study on semi-hollow nano-discs, \textit{i.e.} nanorings in which the central hole goes not through the entire height. 
Such geometry allows to create a bianisotropic directional scattering, which is dependent on the illumination direction.
In other words, light is scattered purely in forward direction, or not -- depending on its incidence direction.
These nanostructures could be used to design reflective metasurfaces, which is only possible thanks to the anisotropy of the individual metasurface unit-cells, inducing a \(2\pi\) phase shift in the reflected amplitude. 
Off resonance, the dielectric metasurface is transparent, which would be impossible using metallic mirrors, due to their high ohmic losses \cite{alaee_all-dielectric_2015}.

\subsection{Applications of Nanoscale Directional Scattering}

The last part of this section is dedicated to an overview on some useful applications of directional scattering by nanoscale particles.

We previously discussed the example of a dielectric metasurface as a lossless mirror, which can be also designed to offer an incidence-dependent reflectivity \cite{alaee_all-dielectric_2015}.
Other all-dielectric metasurfaces have been proposed, offering a generalized Brewster effect for arbitrary angles and wavelengths \cite{paniagua-dominguez_generalized_2016}.
Such metasurfaces are particularly promising for applications in photovoltaics: Light-trapping could significantly reduce reflective losses and increase the efficiency of state-of-the-art solar cells \cite{muskens_design_2008, spinelli_light_2014, priolo_silicon_2014, li_method_2015, mann_opportunities_2016}.

The possibility to render a nanowire completely  invisible \cite{mirzaei_all-dielectric_2015, liu_invisible_2015} has been proposed as a tool to design invisible electric circuits \cite{fan_invisible_2012}.

Directional scattering of the radiation from quantum emitters by dielectric nano-particles has been theoretically predicted \cite{rolly_boosting_2012, rolly_controllable_2013, yang_controlling_2015} and recently also experimentally demonstrated \cite{peter_directional_2017, cihan_silicon_2017}.
Together with the Purcell effect -- the decay rate enhancement for emitters in the proximity of nano-structures \cite{purcell_spontaneous_1946} -- the directionality in the emission renders dielectric particles and nanowires very promising for many applications in field enhanced spectroscopy.
However, the position of the emitter with respect to the particle is crucial for the scattering directionality.
This sensitivity to the emitter location holds generally, for quantum emitters outside \cite{rolly_crucial_2011} as well as for such embedded inside dielectric nano-particles \cite{rocco_controlling_2017}.
On the other hand, this supposed drawback can be actually an advantage and might be used in far-field measurements to gain information about the emitter location on a subwavelength scale.
Using Kerker-type scattering effects, the emission from a nanostructure itself, such as photoluminescence \cite{paniagua-dominguez_enhanced_2013, ramezani_hybrid_2015, brenny_directional_2016} or nonlinear effects like second harmonic generation  \cite{carletti_shaping_2016, xiong_compact_2016}, can be focused into a preferred direction, which is useful in the detection of weak signals from individual nanostructures.

\section{Mie Theory}

In this section, we will explain Kerker-type directional scattering in the context of Mie theory for nanospheres and adapt the same idea in a slightly modified form to infinitely long cylinders (\textit{i.e.} nanowires).
The results are mainly excerpted from an earlier publication. 
Therefore, for more details see also~\cite{wiecha_strongly_2017}.

Mie theory provides an analytical description of the response of spherical or cylindrical particles (with an infinitely long axis in the latter case) to an incident optical field.
The far-field response to an external illumination is written as a multipole series whose coefficients -- the ``Mie scattering coefficients'' \(a_i\) and \(b_i\) -- can be regarded as weights for corresponding electric (\(a_i\)) and magnetic (\(b_i\)) multipole moments.
The expressions ``electric'' and ``magnetic'' also refer to the fact that the magnetic, respectively electric field components, are zero in the scattering plane.

As a short remark we would like to mention that it has been demonstrated, that the Mie coefficients for a cylinder can be written equivalently as a series of Fano profiles, which is explained by the interference between the continuum of modes represented by the incident field and a (sharper) resonant mode (a Mie resonance of the cylinder) \cite{rybin_mie_2013}.
Even though this is not directly linked to the demonstration below, it is an interesting interpretation of Mie resonances, adding a different viewpoint on them in general.

\subsection{Directional Scattering from Spheres and Cylinders}

\subsubsection{Nanospheres}

In the description of the light scattering in spherical particles (which is the ``classical'' Mie theory), all fields are expanded in vector spherical harmonics, leading to a kind of multipole development.
Scattering to the far-field can be written using the scattering amplitude matrix (S-matrix), which connects the incident field \(\mathbf{E}_i\) with the scattered field \(\mathbf{E}_s\) \cite{bohren_absorption_1998}
\begin{equation}\label{eq:scattering_Smatrix_sphere}
 \begin{bmatrix}
  E_{s, \parallel} \\
  E_{s, \perp}
 \end{bmatrix}
  =
 \frac{e^{ - \mathrm{i} k (R - z)}}{ikR}
 \begin{bmatrix}
  S_2 & 0    \\
  0   & S_1
 \end{bmatrix}
 \begin{bmatrix}
  E_{i, \parallel} \\
  E_{i, \perp}
 \end{bmatrix} \, .
\end{equation}
\(k=2\pi/\lambda\) is the wavenumber, \(e^{\mathrm{-i}kR}/(ikR)\) the scattered (outgoing) wave with \(R\) the distance to the sphere center and \(e^{\mathrm{i} k z}\) is the incident plane wave.
For simplicity, we will now assume that the sphere is sufficiently small, such that only the first order of the development has a significant magnitude.
In this case the two nonzero S-matrix elements write \cite{bohren_absorption_1998 ,hosemann_computation_1971}
\begin{equation}\label{eq:Smatrix_S1_S2}
 \begin{aligned}
  S_1 & = \frac{3}{2} \Big( a_1 + b_1 \cos(\varphi) \Big) \, , \\
  S_2 & = \frac{3}{2} \Big( a_1  \cos(\varphi) + b_1 \Big) \, .
 \end{aligned}
\end{equation}
In Eq.~\eqref{eq:Smatrix_S1_S2}, \(\varphi\) is the scattering angle with respect to the incident wave vector, and \(\varphi = 0\) corresponds to the forward direction.
This directly leads to the well-known conditions for exclusive BW or FW scattering:
\begin{equation}\label{eq:pure_FW_BW_conditions_sphere}
 \begin{aligned}
    S_i\Big|_{\varphi=0} & \propto & \Big( a_1 + b_1 \Big) = 0 \quad \text{for pure BW scattering  (2. Kerker),}\\
    S_i\Big|_{\varphi=\pi} & \propto & \pm \Big( a_1 - b_1 \Big) = 0 \quad \text{for pure FW scattering (1. Kerker)}.
 \end{aligned}
\end{equation}
%

\begin{figure}[t]
  \centering
  \includegraphics*{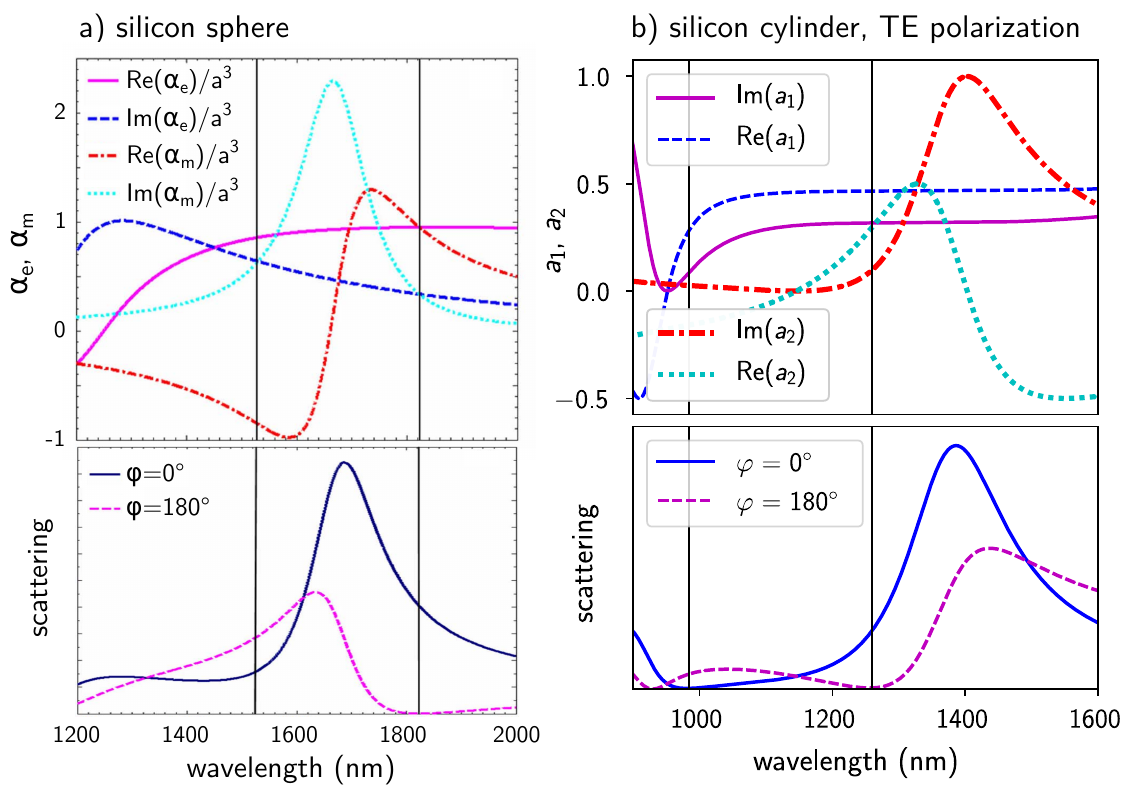}
  \caption[Mie coefficients for a silicon nano-sphere and nano-wire]{
  a) electric and magnetic polarizability for a silicon sphere (\(\epsilon=12\)) of radius \(R=230\,\)nm (top) and directional scattering intensity (\(\varphi = 0^{\circ}, 180^{\circ}\) correspond to FW, BW direction). 
  Adapted with permission from \cite{nieto-vesperinas_angle-suppressed_2011}. Copyright OSA, 2011. Note that the quantities in the top plot of (a) are proportional to the first order (spherical) Mie coefficients: \(\alpha_e \propto \text{i}a_1\) and \(\alpha_m \propto \text{i}b_1\) \cite{garcia-etxarri_strong_2011}.
  b) first and second order electric (TE) Mie coefficients \(a_1\) and \(a_2\) for a silicon nanowire with \(R=230\,\)nm. 
  Bottom: Scattered intensity in FW (\(\varphi = 0^{\circ}\)) and BW (\(\varphi = 180^{\circ}\)) direction.
  Colors are chosen for direct comparison with a).
  The vertical black lines indicate a maximum in BW/FW (left line) and FW/BW (right line) scattering ratio, hence positions where the second, respectively first Kerker's condition are (approximately) satisfied
  } 
  \label{fig:mie_sphere_cylinder}
\end{figure}

\subsubsection{Nanowires}

Infinitely long cylinders can also be treated using Mie theory by developing the fields in vector \textit{cylindrical} harmonics. 
Under normal incidence, the Mie S-matrix writes in this case \cite{bohren_absorption_1998}
\begin{equation}\label{eq:scattering_Smatrix}
 \begin{bmatrix}
  E_{s, \text{TM}} \\
  E_{s, \text{TE}}
 \end{bmatrix}
  = 
e^{\mathrm{i} 3\pi/4} \, \sqrt{\dfrac{2}{\pi k R}} \, e^{\mathrm{i} k R}
 \begin{bmatrix}
  T_1 & 0    \\
  0   & T_2
 \end{bmatrix}
 \begin{bmatrix}
  E_{i, \text{TM}} \\
  E_{i, \text{TE}} \, .
 \end{bmatrix}
\end{equation}
\(k=2\pi/\lambda\) is the wavenumber and \(R\) the distance to the cylinder axis.
As for nanospheres, we assume nanowires of sufficiently small diameter, such that only the lowest two orders significantly contribute to scattering, leading to
\begin{equation}\label{eq:Smatrix_T1_T2}
 \begin{aligned}
  T_1 & = b_1 + 2 b_2 \cos(\varphi) \\
  T_2 & = a_1 + 2 a_2 \cos(\varphi). 
 \end{aligned}
\end{equation}
Again, \(\varphi\) is the scattering angle with respect to the incident wave vector with \(\varphi = 0\) the forward direction.

Under normal incidence, the transverse magnetic (TM) polarized components of the scattered field  are proportional to the S-matrix component \(T_1\) (respectively, the transverse electric (TE) components are proportional to \(T_2\)).
According to Eqs.~\eqref{eq:Smatrix_T1_T2}, scattering from a TE polarized normally incident plane wave (\(\mathbf{E} \perp\) NW axis) is only due to the ``electric'' multipole contributions~\(a_i\).
On the other hand, a TM polarized illumination (\(\mathbf{E} \parallel\) NW axis) induces scattering exclusively via the ``magnetic'' Mie terms~\(b_i\).

In perfect analogy to equation~\eqref{eq:pure_FW_BW_conditions_sphere} we find
\begin{equation}\label{eq:pure_FW_BW_conditions_TM}
 \begin{aligned}
    T_1\Big|_{\varphi=0} & = b_1 + 2 b_2 = 0 \quad\quad \text{for pure BW scattering, and}\\
    T_1\Big|_{\varphi=\pi} & = b_1 - 2 b_2 = 0 \quad\quad \text{for pure FW scattering.}
 \end{aligned}
\end{equation}
The same conditions hold for TE polarization with \(T_2\) and the ``electric'' coefficients \(a_i\). The main difference to the case of a sphere is that we do not have an interference of dipolar electric and dipolar magnetic modes anymore, but an interference between dipolar and quadrupolar modes of the same, ``electric'' or ``magnetic'', character, depending on the incident polarization.
In the upper panels of figure~\ref{fig:mie_sphere_cylinder}, the Mie coefficients \(a_1\) and \(b_1\) for a silicon sphere (Fig.~\ref{fig:mie_sphere_cylinder}a) are compared to the coefficients \(a_1\) and \(a_2\) of a silicon nanowire as function of the wavelength.
Indeed, the coefficients show very similar spectral dependencies, which holds equally for their amplitude and phase.
Comparing the FW/BW scattering spectra of both geometries (bottom plots in figure~\ref{fig:mie_sphere_cylinder}), the similarity in Mie coefficients is reproduced.
Please note also the Fano-like profiles in the scattering resonances, shown in figure~\ref{fig:mie_sphere_cylinder}: 
The interference of a spectrally narrow mode (electric mode for the sphere, ``\(a_2\)''-mode in the case of the nanowire) with a spectrally large ``background-mode'' (sphere: magnetic mode, nanowire: ``\(a_1\)''-mode) leads to the directional response and a Fano-like line shape.

We conclude that interference of multiple spectrally overlapping orders of either electric or magnetic modes in dielectric nanowires can lead to similar FW/BW scattering phenomena as interfering electric and magnetic dipole resonances in dielectric nano-spheres. Different behaviors, hence different directions of light scattering, could be addressed by incident light polarization.

\begin{figure}[t]
  \centering
  \includegraphics*{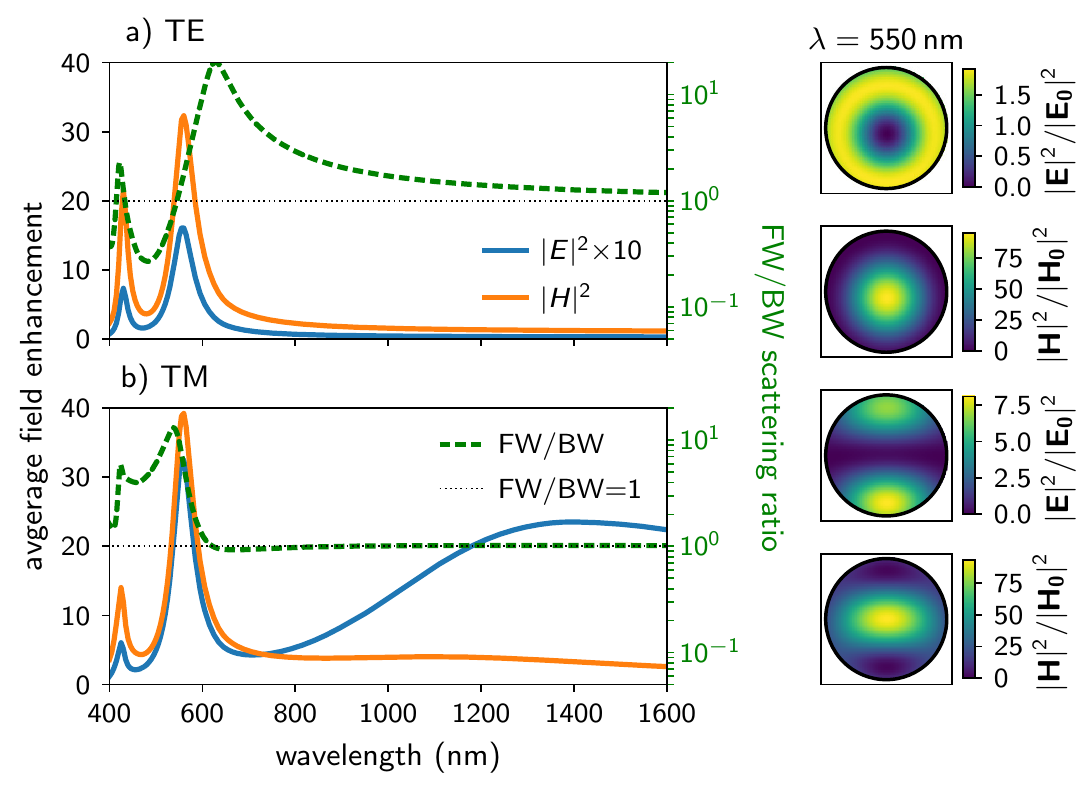}
  \caption[Electric / magnetic field enhancement within \(R=50\,\)nm silicon nanowire]{
  Spectra of the average electric (blue lines) and magnetic (orange lines) field intensity enhancement inside a silicon nanowire of radius \(R=50\,\)nm. 
  Incident plane wave polarized a) TE and b) TM.
  For comparison, the FW/BW ratio of far-field scattering is also shown (dashed green lines, logscale).
  Right: internal field intensity distributions at \(\lambda=550\,\)nm (around the TE\(_{01}\) / TM\(_{11}\) mode). Top: electric, bottom: magnetic field enhancement. 
  Incidence from the top.
  The fields are calculated using Mie theory and normalized to the illumination field intensity.
  } 
  \label{fig:e_b_enhancement_sinw}
\end{figure}

\subsection{Nanowires: Resonant Enhancement of the Electric and Magnetic Field}

It is possible to expand the electromagnetic fields inside the cylinder in the same way as the scattered near- or far-field.
For details, please refer to the textbook of Bohren and Huffmann \cite{bohren_absorption_1998}.
In figure~\ref{fig:e_b_enhancement_sinw} we show the average field intensity enhancement inside a silicon NW (SiNW) of radius \(R=50\,\)nm for illumination with (a)~TE and (b)~TM polarized plane waves, respectively.
The silicon dispersion is taken from the book of Palik \cite{palik_silicon_1997}.
At resonance,  we observe for both the TE and for TM cases not only high electric field intensities (blue lines), but also a strong enhancement of the magnetic field (orange lines).
The magnetic field increases even far stronger compared to the electric field intensity, an observation which is in agreement with recent experimental results from dielectric cylinders in the GHz regime \cite{kapitanova_giant_2017}.
We conclude that a simultaneous excitation of strong electric and magnetic fields occurs in dielectric, non-magnetic (\textit{i.e.} \(\mu_r = 1\)) NWs, even under pure TE or TM polarized illumination and normal incidence.
Hence, in analogy to the findings of Kerker \textit{et al.}, the observed directionality (dashed green lines in Fig.~\ref{fig:e_b_enhancement_sinw}) can be interpreted as a result of the interference between \textit{``effective'' electric and magnetic modes}.
In particular, at the non-degenerate, fundamental TM\(_{01}\) mode (\(\lambda \approx 1400\,\)nm), where only the internal \textit{electric} field shows a resonant enhancement while the \textit{magnetic} field intensity follows a flat line beyond \(\lambda \gtrsim 700\,\)nm (see Fig.~\ref{fig:e_b_enhancement_sinw}b), no directional scattering is obtained. This is also the case for any small diameter nanowire supporting the fundamental resonant mode only.
For illustration the electric and magnetic field intensity patterns inside the NW are shown on the right of Fig.~\ref{fig:e_b_enhancement_sinw} at~\(\lambda=550\,\)nm.

\begin{figure}[t]
  \centering
  \includegraphics*{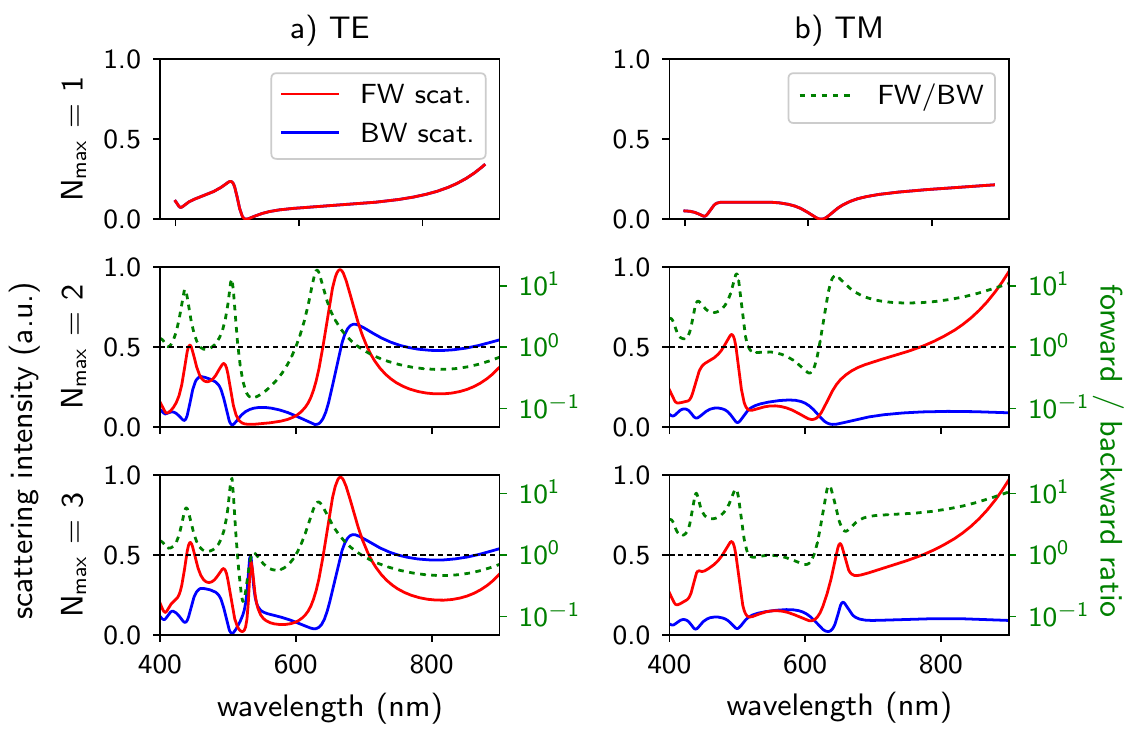}
  \caption[Contributing Mie multipole orders to directional scattering in SiNWs]{
  Mie development of the FW/BW scattering from a SiNW with radius \(R=100\,\)nm for the first 3 Mie coefficients (TE: \(a_n\), TM: \(b_n\)). 
  a) TE, b) TM polarized incident plane wave.
  \(N_{\text{max}}\) corresponds to the number of Mie terms: \(n = 0\), \(n\in\{0,1\}\), \(n\in\{0,1,2\}\) (from top to bottom). 
  TE/TM data is normalized separately.
  FW/BW ratios (green dashed lines) are plotted on a logscale
  } 
  \label{fig:sinw_dirscat_required_orders}
\end{figure}

\subsection{Nanowires: Multipolar Contributions to Directional Scattering }

Previously, we used only the two first coefficients of the field expansion. We want to assess if this approximation remains valid for larger NWs, and how many orders of multipole contributions are necessary to describe with a sufficient agreement the directional scattering phenomena.
Let us illustrate this empirically by taking a SiNW of radius~\(R=100\,\)nm for example.
Since directional scattering is a result of the interference between multiple simultaneously excited modes, the FW/BW resolved scattered intensity cannot be plotted individually for the different contributing scattering coefficients~\(a_n\) and~\(b_n\) (for TE and TM polarized, normal incidence, respectively). 
In figure~\ref{fig:sinw_dirscat_required_orders}, the FW and BW scattered intensity from a normally illuminated SiNW (\(R=100\,\)nm) is calculated successively for an increasing number of contributing terms.
Figure~\ref{fig:sinw_dirscat_required_orders}a shows the scattering under TE polarized illumination, Fig.~\ref{fig:sinw_dirscat_required_orders}b the TM case.
We have shown elsewhere \cite{wiecha_strongly_2017} that for a NW of~\(R=50\,\)nm radius, terms higher than the first two orders are negligible. 
Yet, even for a larger NW as shown in Fig.~\ref{fig:sinw_dirscat_required_orders}, the response is mostly determined by the first two orders of the Mie series, while only few additional features arise if third order modes are considered as well. 
In conclusion, despite some missing spectral features, using only the first two Mie orders gives already a very good approximation even in the case of large dielectric NWs.

\section{Directional Scattering from Silicon Nanowires}

In this section, we present experiments performed on silicon nanowires showing the directional scattering phenomenon as function of size and shape. 
We compare the experimental data to Mie theory and numerical simulations, confirming the accuracy of the predictions.
Having confirmed the validity of our approach, we analyze theoretically a more sophisticated system of two normally illuminated parallel nanowires, lying on a plane. 
We find that such a system can be used to switch the scattering direction between forward and backward, simply by changing the distance between the two wires.

\begin{figure}[t]
  \centering
  \includegraphics*{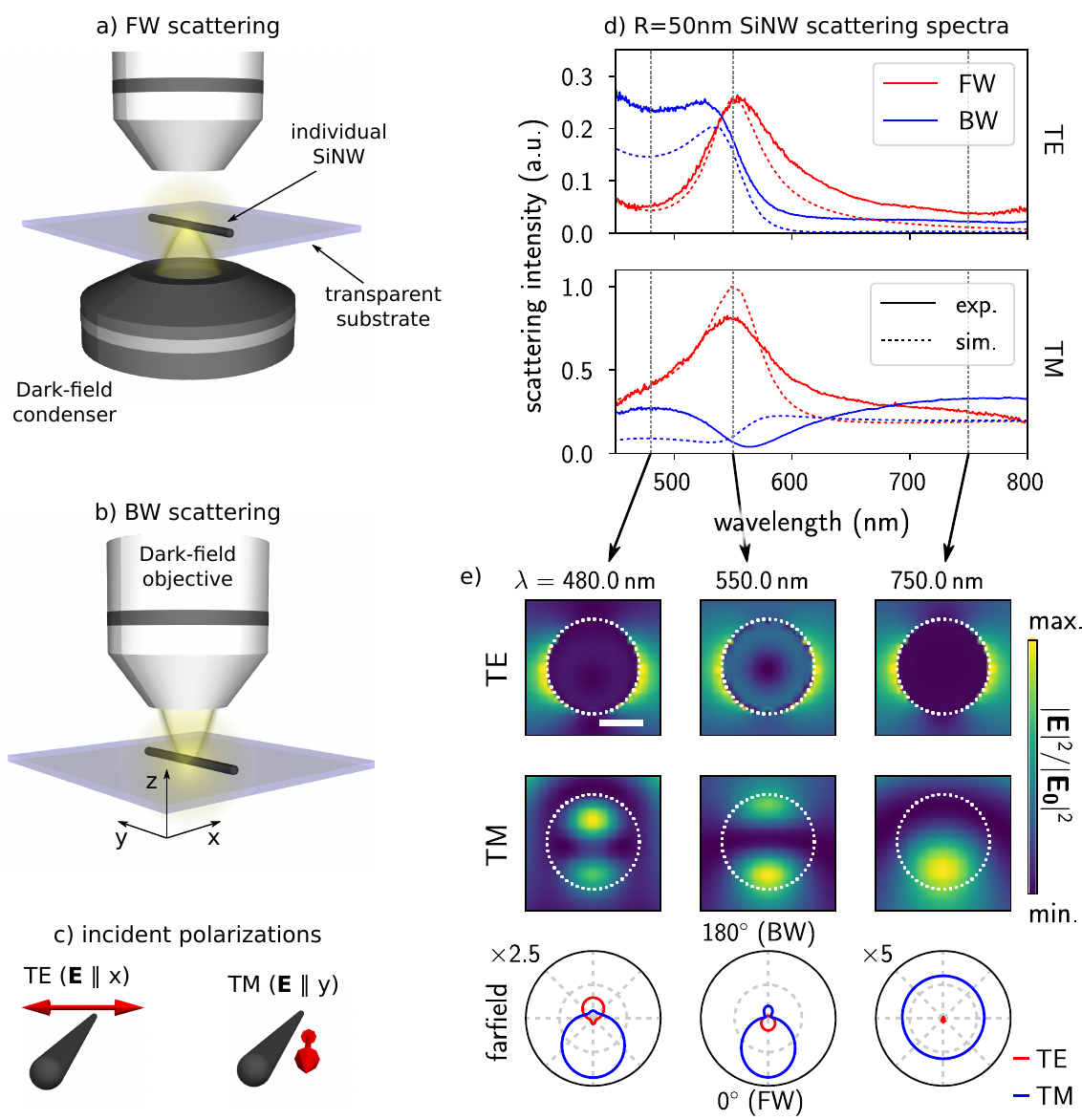}
  \caption[Experimental setup; directional scattering from cylindrical SiNWs]{
  Sketch of the experimental setup for (a) forward scattering and (b) backward scattering measurements. 
  (c) Incident polarization configurations. The electric field is orientated either perpendicular (TE) or parallel (TM) to the NW axis.
  (d) Experimental (solid lines) and simulated (dashed lines) FW/BW scattering spectra (red/blue) for a cylindrical silicon NW of diameter \(R\approx 50\,\)nm.
  Top and bottom plot show the case of TE and TM polarized incident plane waves, respectively.
  (e) Mie-calculated nearfield (top row: TE, center row: TM incidence) and farfield patterns (bottom row; TE/TM: red/blue) for a \(R = 50\,\)nm SiNW diameter at selected wavelengths, indicated by dashed vertical lines and arrows in (d). 
  Plane wave incident from the top. Scale bar is \(50\,\)nm.
  } 
  \label{fig:cylindrical_sinws}
\end{figure}

\subsection{Cylindrical Nanowires}

We start with the simplest possible geometry: A cylindrical SiNW in vacuum.
The SiNWs are VLS grown~\cite{dhalluin_silicon_2010} and dispersed on a transparent silica substrate with lithographic markers, so the exactly same NW can be examined on different experimental setups.
We perform standard dark field microscopy either in reflection (BW scattering) or in transmission geometry (FW scattering). 
The measurement setup is schematically shown in Fig.~\ref{fig:cylindrical_sinws}a and \ref{fig:cylindrical_sinws}b for FW and BW scattering, respectively. 
Details on the technique are described elsewhere~\cite{wiecha_strongly_2017}.
The incident light is polarized either perpendicular (TE) or along the NW axis (TM), as illustrated in Fig.~\ref{fig:cylindrical_sinws}c.
We use the scattering from small cylindrical nanowires (\(R\approx 25\,\)nm) for normalization of the FW and BW spectra:
In sufficiently small NWs only the fundamental dipolar TM\(_{01}\) mode is excited, which results in an omni-directional scattering (corresponding to a dipolar source along the NW axis, as explained in the previous section).
We hence assume that the FW and BW scattered intensities are of equal strength and normalize all spectra using this reference.
We compare our experimental results to Mie theory and 2D~simulations by the Green dyadic method (``GDM'', assuming structures of infinite length along \(Y\) \cite{martin_generalized_1995, paulus_greens_2001}).
The GDM simulations reproduce Mie with almost perfect agreement (see also~\cite{wiecha_strongly_2017}).

Results of the scattering experiments on a SiNW of radius~\(R\approx 50\,\)nm are shown in Fig.~\ref{fig:cylindrical_sinws}d for TE (top) and TM (bottom) illumination. 
In the case of TE polarized illumination FW scattering occurs for large wavelengths (\(\lambda \gtrsim 550\,\)nm), while BW scattering takes over at shorter wavelengths (\(\lambda \lesssim 550\,\)nm).
In the TM case on the other hand, FW scattering dominates over the whole accessible spectral range, however with a pronounced peak around \(550\,\)nm, where a maximum of scattered FW intensity coincides with a minimum in BW scattering.
Near-field plots and corresponding far-field scattering patterns are shown in figure~\ref{fig:cylindrical_sinws}e (top and bottom, respectively) for selected wavelengths, indicated by dashed black lines in figure~\ref{fig:cylindrical_sinws}d.
Note that around \(480\,\)nm, we obtain the possibility to invert the main scattering direction by simply flipping the polarization from TM to TE (see also Fig.~\ref{fig:cylindrical_sinws}e, bottom left). Via the NW diameter, this spectral zone can be tuned to other frequencies.
At longer wavelengths (\(\lambda\gtrsim 700\,\)nm), only the nondegenerate TM\(_{01}\) mode exists, leading to very weak overall scattering under TE incidence and to the aforementioned, omnidirectional radiation pattern in the TM geometry.
Interestingly, while it is possible to induce directional BW scattering in the TE configuration, under TM illumination BW scattering is generally very weak and mainly FW scattering occurs -- with the exception of a uniform scattering when only the TM\(_{01}\) mode is excited.

In summary, we note that although we do observe BW scattering (mainly under TE polarization) it mostly remains weak compared to the FW scattered light. 
In the case of TM polarized illumination, the FW/BW scattering ratio is even almost exclusively \(\gtrsim 1\).

\begin{figure}[t]
  \centering
  \includegraphics*{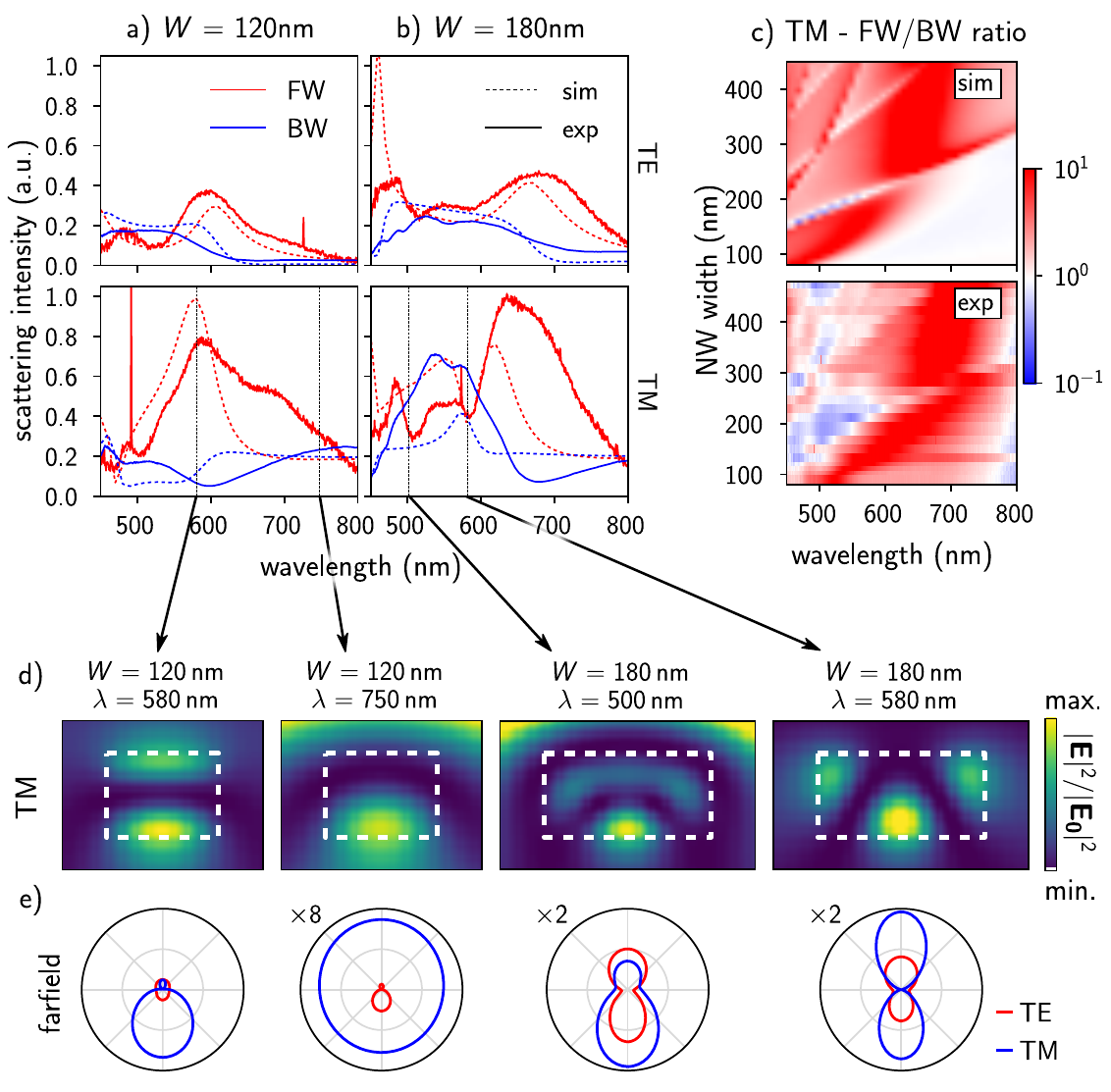}
  \caption[Directional scattering from rectangular SiNWs]{
  (a-b) Scattering spectra of rectangular shaped SiNWs of a fixed length and height \(L=7\,\)\textmu m and \(H=90\,\)nm, with widths of a) \(W=120\,\)nm and b) \(W-180\,\)nm. 
  Plane wave illumination, polarized perpendicular (TE, top) or along the NW axis (TM, bottom).
  Red lines indicate forward, blue lines backward scattering. 
  Dashed and solid lines represent simulated and experimental data, respectively.
  c) simulated (top) and measured (bottom) FW/BW scattering ratio for TM polarization using a logarithmic color scale.
  d) Nearfield intensity distribution in and around the SiNWs at selected wavelengths (indicated by vertical lines in (a-b) and by arrows) under TM polarized illumination (incident from the top). 
  Dashed white lines illustrate the NW shape.
  e) Farfield scattering pattern for the corresponding cases shown in d).
  TM: blue, TE: red lines
  } 
  \label{fig:rectangular_sinws}
\end{figure}

\begin{figure}[t]
  \centering
  \includegraphics*{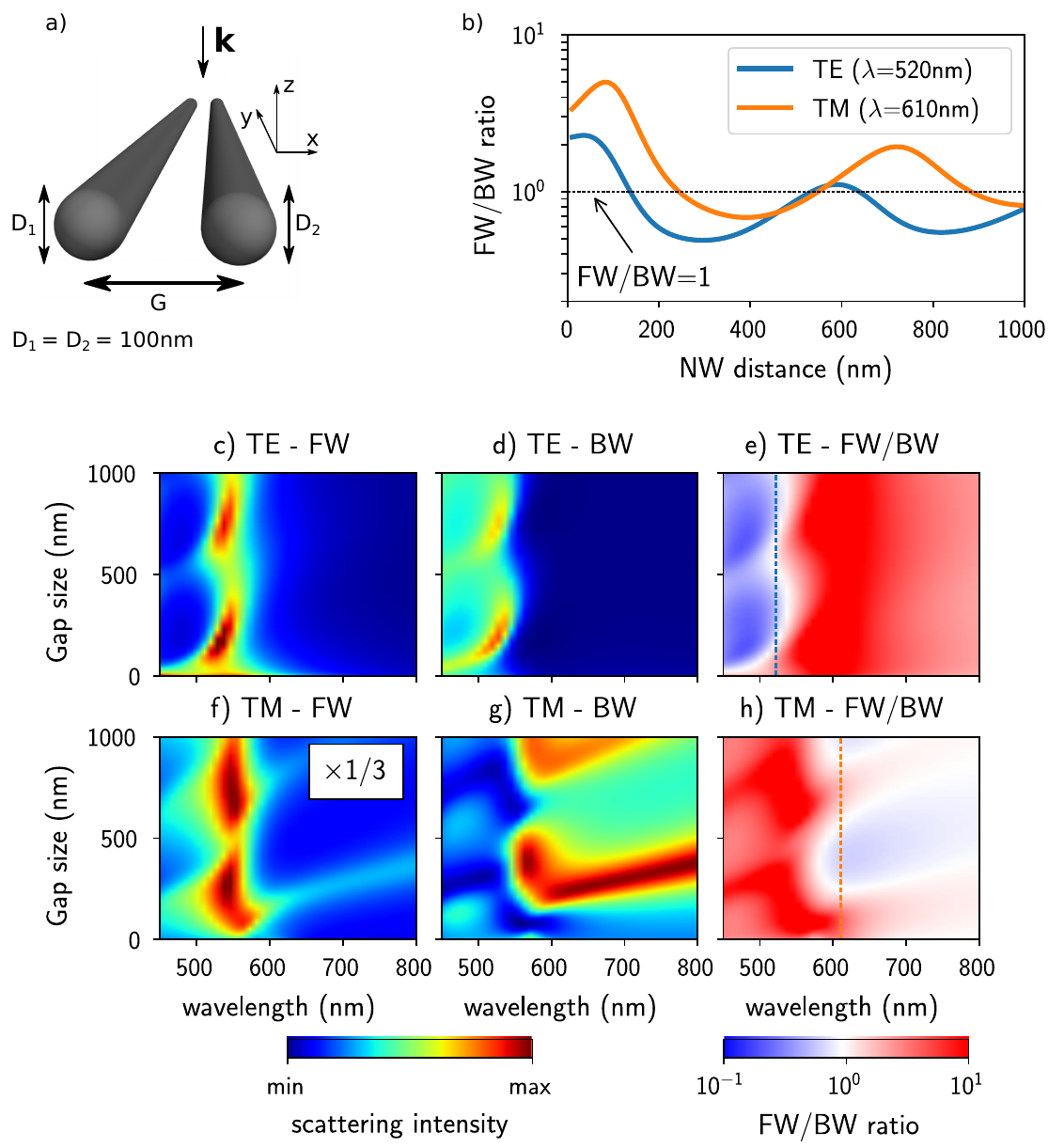}
  \caption[Directional scattering from coupled SiNWs]{
  a) Sketch of the geometry: Two parallel silicon NWs in the \(XY\), of equal diameter \(D_1=D_2=100\,\)nm (\(\rightarrow R=50\,\)nm) and separated by a ``gap'' \(G\), are normally illuminated (along \(\mathbf{\hat e}_z\)) by a plane wave, polarized either along the mutual long axis direction (``TM'') or perpendicular to it (``TE'').
  b) Simulated forward to backward scattering ratio as function of the distance \(G\) between the two NWs for two TE and TM at fixed wavelengths: \(\lambda_{\text{TE}}=520\,\)nm and \(\lambda_{\text{TM}}=610\,\)nm.
  (c-h) show the simulated FW/BW resolved scattering spectra of the coupled NWs as function of the gap \(G\).
  (c-e) show forward scattering, backward scattering and their ratio for TE polarization, (f-h) similar but for TM polarization.
  Vertical dashed lines in e) and h) indicate the spectral positions of the profiles shown in b).
  The FW/BW ratios in b), e) and h) are shown on a logarithmic scale
  } 
  \label{fig:coupled_sinws}
\end{figure}

\subsection{Rectangular Nanowires}

In a second step we analyze what happens if the cylindrical symmetry of the nanowire cross section is broken. 
We therefore fabricate SiNWs of rectangular section by electron beam lithography (EBL) and subsequent dry-etching \cite{han_realization_2011, guerfi_high_2013} on a silicon-on-quartz (SOQ) substrate \cite{moriceau_materials_2014}. 
A great advantage of our top-down approach on SOQ is the possibility to create  silicon nanostructures of arbitrary shape by EBL on a transparent substrate from single crystalline silicon.
Using this material rather than deposited polycrystalline or amorphous silicon guarantees that the best optical properties of single crystal silicon are kept. 
Defined by the thickness of the silicon layer on the SOQ substrate (\(H=90\,\)nm), the height of the rectangular nanowires is constant. The rectangular section is varied by changing the SiNWs width.
The length \(L=7\,\)\textmu m is chosen large compared to the focal spot of the illuminating optics in order to obtain a purely Mie-like response \cite{traviss_antenna_2015}.
All NWs have excellent surface properties, low roughness and steep flanks, verified by scanning electron microscopy.

The results of our systematic FW/BW scattering measurements are shown in figure~\ref{fig:rectangular_sinws}c (for TM polarization).
Selected spectra for SiNWs of width \(W=120\,\)nm and \(W=180\,\)nm are shown in figure~\ref{fig:rectangular_sinws}a-b, respectively, where TE polarized illumination is shown in the top, the TM case in the bottom plot.
The comparison of experiments with GDM simulations shows a very good agreement.
Having a look at the FW/BW ratios (Fig.~\ref{fig:rectangular_sinws}c, TM polarization) we observe that mostly forward scattering occurs.
This is similar to our observations on cylindrical SiNWs.
Figure~\ref{fig:rectangular_sinws}d and e) finally show selected simulated near-field plots and far-field scattering patterns for TM polarized illumination.
In particular, the first two panels in Fig.~\ref{fig:rectangular_sinws}d-e show, for a NW of nearly symmetric section, the excitation of the first strongly directional mode (corresponding to TE\(_{01}\)/TM\(_{11}\)) and the region of omnidirectional scattering in the red spectrum (corresponding to TE\(_{01}\)/TM\(_{11}\)), which are almost identically found as in cylindrical NWs.

Under TM excitation, we observe branches of a kind of Fano resonance, which do not occur in cylindrical NWs: 
In narrow spectral zones, the FW/BW ratio is almost unity (see Fig.~\ref{fig:rectangular_sinws}b around \(\lambda = 580\,\)nm and Fig.~\ref{fig:rectangular_sinws}c, white diagonal branches).
In a confined spectral window the otherwise strong forward scattering is suddenly suppressed, while BW scattering increases. 
Resonances with such field profiles are observed neither under TE polarization, nor in symmetric SiNWs (see also~\cite{wiecha_strongly_2017}).
These sharp features are a result of horizontal guided modes along the SiNW width (\(X\)-direction, in the NW cross-sectional plane) and their interference with the leaky ``background'' mode of, in comparison, large spectral width.
The wire side facets act as the Fabry-Perot cavity mirrors and the nanowire slab
has an effective index of \(n_{\text{eff}}=3.45\) (\(n_{\text{Si}}\approx 4.0\)) for the supported guided mode in the case of TM polarized illumination at \(\lambda=580\,\)nm. 
With this effective index, a standing wave pattern with three lobes matches perfectly a nanowire width of \(180\,\)nm, in nice agreement to the wavelength and width at which the resonance occurs. 
The calculated near-field pattern (very right panel of figure~\ref{fig:rectangular_sinws}d) finally confirms the presence of a guided mode. 
We verified the guided mode assumption also for the other branches in figure~\ref{fig:rectangular_sinws}c, at each of which the NW width corresponds to an odd integer multiple of half the wavelength divided by the effective index.
A standing-wave pattern of increased order can be found at these positions in the near-field plots.
We note that former studies on scattering from rectangular dielectric NWs used Fabry-Perot modes, reflected between the side-walls of the wire, for the analytical description of the scattering \cite{fan_optical_2014, landreman_fabry-perot_2016}.

\subsection{Coupled Nanowires}

Finally we analyze a system composed of two identical, parallel silicon nanowires  illuminated normally with respect to the plane defined by their axes, as schematically shown in Fig.~\ref{fig:coupled_sinws}a.
The radius of the wires is \(R=50\,\)nm, (\textit{c.f.} also figure~\ref{fig:cylindrical_sinws}), the incident plane wave is polarized either perpendicular (TE) or parallel to the wire axis (TM).
Using the GDM, the scattered intensity is calculated in forward and backward direction. The results are shown in Fig.~\ref{fig:coupled_sinws} as function of the incident wavelength and polarization as well as the distance between the NWs, also labeled ``gap'' \(G\).

Figs.~\ref{fig:coupled_sinws}c-e show c) forward, d) backward scattering and e) the FW/BW ratio in the case of TE polarization.
Figs.~\ref{fig:coupled_sinws}f-h show the same for the TM case.
We find a strong modulation of the scattering, which is in agreement with former observations in the total scattering signal \cite{mirzaei_electric_2015}.
Interestingly, not only the scattering intensity itself shows a modulation of high contrast; likewise, the FW/BW ratio shows a significant dependence on the distance between the two SiNWs.
Figure~\ref{fig:coupled_sinws}b shows the FW/BW scattering ratio for TE (blue dashed line) and TM polarization (orange dashed line) at two fixed wavelengths of particularly strong contrast in the scattering signal.
Obviously, at selected wavelengths, it is possible to toggle between mainly forward, isotropic, and mainly backward scattering simply by varying the gap size between the two nanowires. Furthermore, an interesting feature of coupled nanowires is to increase the scattering efficiency either in forward or backward direction, compared to the case of a single nanowire where it is rather low at Kerker's conditions \cite{shibanuma_unidirectional_2016}. This is due to the fact that Kerker's conditions are usually fulfilled out of a resonance.
This observation opens perspectives for several applications in sensing or field-enhanced spectroscopy by realizing more efficient directional dielectric nanoantennas. 
We could for instance imagine two parallel dielectric nanowires embedded in a flexible transparent matrix for sub-wavelength optical distance measurements at visible frequencies.

\section*{Conclusion and Outlook}

We compared plasmonics with the emerging field of high-index dielectric nanostructures and gave an overview about the recent research in nano-optics on directional scattering at the single particle level.
We introduced the Kerker's conditions, their representation for nonmagnetic materials at visible frequencies, and put them in the context of Fano resonances in general.
Furthermore, we compared the conditions for unidirectional scattering in Mie theory for spherical particles and cylindrical nanowires.
We experimentally demonstrated directional scattering first in the ``Mie case'' of cylindrical silicon nanowires, where we found the possibility to switch between forward and backward scattering by a simple rotation of the incident polarization. 
Subsequently, we studied the effects of an asymmetric nanowire shape.
For rectangular NWs of increasing height/width aspect-ratio, we observed the apparition of Fano-like modes due to the occurrence of spectrally sharp guided modes along the NW width.
Finally, we theoretically studied a system of two coupled SiNWs and found that such an arrangement allows to toggle between FW and BW scattering, simply by adjusting the distance between the coupled wires.
Optical Fano resonances in nanostructures open perspectives for light management and guiding at the nanometer scale, for sub diffraction-limited optical sensing or for applications in field-enhanced spectroscopies

\bibliographystyle{unsrt}
\bibliography{2017_fano_resonances_dir.bib}
%


\printindex
\end{document}